\documentclass[]{aastex631}

\usepackage{amsmath}
\usepackage{mathtools} 

\begin{document}

\title[A unified FRB model]{A unified model of active repeating fast radio bursts associated with persistent radio sources}
%\title[Young magnetars in massive binary embedded in a SNR as sources of FRBs]{Evidence of young magnetars in massive binary embedded in a supernova remnant as sources of active fast radio bursts}

\author[0000-0003-4157-7714]{F. Y. Wang}\thanks{E-mail: fayinwang@nju.edu.cn}
\affiliation{School of Astronomy and Space Science, Nanjing University, Nanjing 210093, China}
\affiliation{Key Laboratory of Modern Astronomy and Astrophysics (Nanjing University), Ministry of Education, Nanjing 210093, China}
\author{H. T. Lan}
\affiliation{School of Astronomy and Space Science, Nanjing University, Nanjing 210093, China}
\author{Z. Y. Zhao}
\affiliation{School of Astronomy and Space Science, Nanjing University, Nanjing 210093, China}
\author{Q. Wu}
\affiliation{School of Astronomy and Space Science, Nanjing University, Nanjing 210093, China}
\author{Y. Feng}
\affiliation{Research Center for Astronomical Computing, Zhejiang Laboratory, Hangzhou 311100, China}
\affiliation{Institute for Astronomy, School of Physics, Zhejiang University, Hangzhou 310027, China}
\author{S. X. Yi}
\affiliation{School of Physics and Physical Engineering, Qufu Normal University, Qufu 273165, China}
\author{Z. G. Dai}\thanks{E-mail: daizg@ustc.edu.cn}
\affiliation{Department of Astronomy, University of Science and Technology of China, Hefei 230026, China}
\author{K. S. Cheng}
\affiliation{Department of Physics, The University of Hong Kong, Pokfulam Road, Hong Kong, China}

\begin{abstract}
Fast radio bursts (FRBs) are intense pulses with unknown origins. A subclass of repeating FRBs show some common features, such as associated compact persistent radio sources (PRSs), high burst rates, and large host-galaxy dispersion measures (DMs). Meanwhile, they show diverse DM and rotation measure (RM) variations, which cannot be explained by current models. A unified model urgently needs to be established. %Many models based on young magnetars have been proposed to account for these features. However, a unified model has not yet been established. 
Here we show the first evidence for a supernova remnant surrounding the FRB 20190520B source. We then demonstrate that the five active repeating FRB sources associated with PRSs can be understood within a single model in which central objects are young magnetars in massive binary systems embedded in supernova remnants. This model naturally predicts distinct variations of DM and RM for repeating FRBs. Crucially, young magnetar wind nebulae can generate bright PRSs. As a magnetar becomes older, the luminosity of a PRS will fade, which can naturally explain less-luminous PRSs for some active FRBs. %Applying this model to another active FRB, it can also reproduce the RM temporal variation. 
Our results support a unified population of active FRBs in dynamic magnetized environments.
%Here, we propose that the sources of these active FRBs are newborn magnetars embedded in supernova remnants (SNRs) in massive binary system. The stellar wind and possible decretion disk of the massive star, together with the SNRs, can account for the distinct variations of DM and RM, such as decrease and increase of DM and RM, and RM reversal. The magnetar wind nebula and synchrotron radiation from shocks produced by collisions between the stellar wind and magnetar wind can generate bright PRSs. This model can naturally explain the luminous PRSs and the distinct variations of DM and RM of active FRBs.
\end{abstract}


\section{Introduction}

Fast radio burts (FRBs) are energetic pulses of coherent radio emission \citep{Lorimer2007,Xiao2021,Zhang2023}. At present, thousands of FRB sources
had been discovered \citep{Wu2024}. They can be divided into two groups, apparent non-repeating FRBs and repeating FRBs \citep{CHIME/FRBCollaboration2021,Chime/FrbCollaboration2026}. Repeating FRBs show diverse
observational properties, including dispersion measure (DM) variation, rotation measure (RM) variation, and association with persistent radio sources (PRSs) \citep{Chatterjee2017,Niu2022,Bruni2024,Bruni2024b,ZhangX2025,Moroianu2025}. 

For the first discovered repeating FRB 20121102A, long-term observation shows that its DM increases during eight years since its discovery \citep{Hessels2019,Oostrum2020,Li2021}, and then declines \citep{WangP2025,Snelders2025}. While the RM value of FRB 20121102A decreases from $10^5~\mathrm{rad\,m^{-2}}$ to $3\times 10^4~\mathrm{rad\,m^{-2}}$ at the year of 2023 \citep{Michilli2018,WangP2025}.
Another active FRB source with extreme RM is FRB 20190520B. Its RM value is about $10^4~\mathrm{rad\,m^{-2}}$ \citep{Niu2022}. It is highly variable and shows twice sign reversal \citep{Anna-Thomas2023}. This FRB source has an extremely large host DM with DM$_{\rm{host}}\sim 900$ pc cm$^{-3}$ \citep{Niu2022}. 
The DM declines with a rate of about 12 pc cm$^{-3}$ year$^{-1}$.
The RM variation of FRB 20201124A is also complex \citep{Xu2022}. It show oscillation in short timescale, followed by a
constant RM phase. After subtracting the RM value of the constant phase, the RM contributed the source vicinity
shows sign reversal \citep{Wang2022}. The DM of this FRB source is stable, with a fluctuation
of about 10 pc cm$^{-3}$ between burst DM \citep{Xu2022}. 

The above three FRB sources are spatially coincident with PRSs. From the very large array interferometric observation, a compact (size$<0.7$ pc) luminous ($\nu L_{\nu}\sim 10^{39}$ erg s$^{-1}$) PRS is coincident to FRB 20121102A \citep{Marcote2017}. The spectrum is flat with a negative slope. FRB 20190520B is also associated with a compact PRS, with similar properties of that of FRB 20121102A \citep{Niu2022,Bhardwaj2023}. Recently,
a potential PRS was reported for FRB 20201124A \citep{Bruni2024}. However, the luminosity is about two orders of magnitude smaller than that of FRB 20121102A, and the spectral index $0.97\pm0.54$ is positive \citep{Bruni2024}. This low luminosity may be due to star formation \citep{Nimmo2022,Dong2024}. 

Other FRB repeaters also show complex and diverse RM and DM variations \citep{Feng2025,Cook2026}. Secular trends of DM and RM for FRB 20180301A are found \citep{Kumar2023}. Combined with the previous observations by \cite{Luo2020}, it shows irregular variation of DM and RM. FRB 20180916B exhibited RM variation with a stochastic component and a linearly increasing trend \citep{Mckinven2023}. Following observations show that it may change to a stochastic trend. From the observations of the FAST and Parkes telescopes for nearly three years, the RM of FRB 20220529A varied slowly between -300 and 300 rad m$^{-2}$ with several sign reversals in the first two years \citep{Li2026}. Then it jumped to about 2000 rad m$^{-2}$ within two months and recovered to the normal value in 14 days, possibly due to a stellar flare from companion. A possible period of 200 days for the RM evolution with significance of 4.2$\sigma$ was reported recently \citep{Liang2025}. The abrupt RM increase and periodic RM evolution strongly support the binary origin of this source. FRB 20190417A also shows large RM variation from about 4000 to 5000 rad m$^{-2}$ \citep{Feng2025,Moroianu2025}. In summary, RM and DM variations are diverse for repeating FRBs.

However, theoretical models cannot naturally account for these variations. For example, after the first localization of FRB 20121102A, \cite{Metzger2017} proposed that it originates from a
young magnetar embedded within a young supernova remnant (SNR). However, theoretical study shows that a young SNR will cause RM monotonic decline, and DM decrease first and then increase \citep{Piro2016,Yang2017,Piro2018,Zhao2021b,Katz2022}. Obviously, the predicted RM and DM variations contradict observations. In order to explain the RM variations of FRB 20201124A and FRB 20190520B, a magentar and massive star binary is proposed \citep{Wang2022,Zhao2023,Rajwade2023,ZhangB2025}. 
The binary model can well reproduce the RM variations of repeating FRBs. However, it cannot cause the DM decreasing with a rate of a few tens per year. This large contribution of DM will lead significant free-free absorption with optical depth $\tau>1$. So radio bursts will be absorbed. 

The phenomenological similarities between these FRB sources invite a unified physical model. Here, we propose that the sources of these active FRBs are young magnetars embedded in SNRs in massive binary systems. %The model is displayed in Figure \ref{fig1}. FRBs are generated by a young magnetar. A stellar wind of the massive companion, together with the SNR, can account for the variations of DM and RM. Luminous PRSs are powered by synchrotron emission from relativistic charged particles of a magnetar wind nebula (MWN), with rotational energy or magnetic energy injection. 
This paper is organized as follows. The model is presented in Section \ref{model}. In Section \ref{sec_result}, we apply our unified model to explain the DM and RM evolution of different FRB sources. Finally, conclusions and discussion are given in section \ref{Conclusion and Discussion}.

\section{The unified model}\label{model}
The model is displayed in Figure \ref{fig1}.
The binary system consists a newborn magnetar and a massive star. The born of magnetar is companion with large mass ejection ($2 M_{\odot}\sim 10 M_{\odot}$).
The ejecta drives a blast wave into the local environment. The
forward and reverse shock wave will be generated as a result of
the interaction between the energetic ejecta and the surrounding
medium. The shocked regions have high temperature and density. Meanwhile, the new-born magnetar can driven relativistic wind through rapid rotation. It interacts with the surrounding medium, generating a luminous magnetar wind nebula (MWN). The magnetic-energy injection from the new-born magnetar can also power a luminous MWN. The luminous MWN can account for the observed PRS of FRBs. 

\subsection{Wind Interactions}\label{sec_wind}
\begin{figure} 
	\centering
	\includegraphics[width=\textwidth]{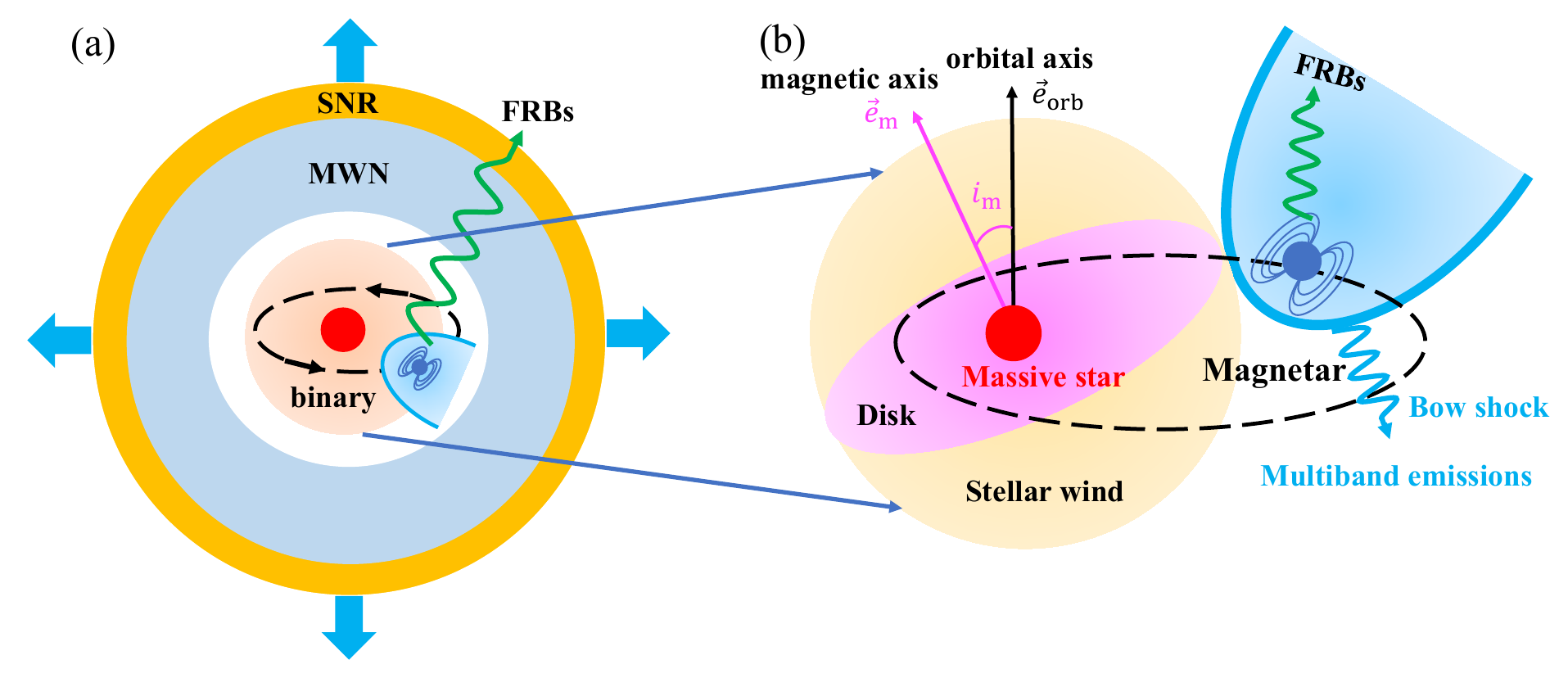}
	\caption{\textbf{A schematic diagram of the unified model.} \textbf{a}, A young magnetar-massive star binary embedded in a magnetar wind nebula and supernova remnant. FRBs are powered by the activity of the magnetar. Radio signals propagate through the stellar wind and supernova ejecta, which can caused DM and RM variations. The magnetar injects magnetic energy or rotational energy into the magnetar wind nebula generating synchrotron radiation, observed as the persistent radio source. \textbf{b}, Geometry of the collision between the stellar wind of massive star companion and the magnetar wind. The bow shock generated by the collision can power multi-band radiation, analog to high-mass gamma-ray binaries. In addition to the stellar wind, some Oe/Be stars can possess an equatorial decretion disk, which can contribute DM and RM.
	}
	\label{fig1}
\end{figure}

%ZZY please add the argument that large DM variation cannot be caused by binary, and may be SNR.
A magnetar/Be star binary is proposed to interpret the DM and RM variations of FRBs \citep{Wang2022,Zhao2023}. Following our model, we here outline the long-term DM evolution from the stellar wind of massive stars. A schematic diagram of the magnetar/massive star binary system is shown in Figure \ref{fig1}. The stellar wind is approximately isotropic, and the electron number density in the stellar wind from the radial distance $r$ is
\begin{equation}\label{eq:nw}
n_{\mathrm{w}}(r)=n_{\mathrm{w}, 0}\left(\frac{r}{R_{\star}}\right)^{-2},
\end{equation}
where $R_{\star}$ is the radius of the massive star. The density at the surface of the massive star is $n_{\mathrm{w}, 0}=\dot{M} / 4 \pi R_{\star}^2 v_{\mathrm{w}} \mu_{\mathrm{i}} m_{\mathrm{p}}$, where $\dot{M}$ is the mass-loss rate, $v_{\mathrm{w}}$ is the wind velocity, $\mu_{\mathrm{i}}\simeq 1.29$ is the mean ion molecular weight \citep{Dubus2013} and $m_{\mathrm{p}}$ is the mass of the protons. The DM variations from the stellar wind is
\begin{equation}\label{eq:dm_wind}
\Delta \mathrm{DM}=\int_{\mathrm{LOS}} n_{\mathrm{e}}(l) \mathrm{d} l.
\end{equation}

From geometry, we know that the distance between any point on the line of sight and the center of the star satisfies
\begin{equation}\label{eq:r}
r^2=d^2+l^2+2 d l\left(\boldsymbol{e}_{\mathrm{mag}} \cdot \boldsymbol{e}_{\mathrm{obs}}\right),
\end{equation}
where $\boldsymbol{e}_{\mathrm{obs}}=\left(\sin i_{\mathrm{o}} \cos \phi_{\mathrm{o}}, \sin \theta_{\mathrm{o}} \sin \phi_{\mathrm{o}}, \cos \theta_{\mathrm{o}}\right)$ is the unit vector of the observer direction, $\boldsymbol{e}_{\mathrm{mag}}=(\cos \phi, \sin \phi, 0)$ and $\phi$ is true anomaly. 
The orbital separation is
\begin{equation}\label{eq:d}
d=\frac{a\left(1-e^2\right)}{1+e \cos \phi},\quad \frac{P_{\mathrm{orb}}^2}{a^3}=\frac{4 \pi^2}{G\left(M_{\star}+m\right)},
\end{equation}
where $e$ is the eccentricity of the orbit, $P_{\mathrm{orb}}$ is the orbital period, $M_{\star}$ is the stellar mass and $m\simeq1.4M_{\odot}$ is the magnetar mass. Combing with Equations (\ref{eq:nw}), (\ref{eq:r}) and (\ref{eq:d}), the integral of Equation (\ref{eq:dm_wind}) can be evaluated analytically
\begin{equation}
\begin{aligned}
\Delta \mathrm{DM}_{\mathrm{w}}(\phi) & =\int_{l_\mathrm{s} }^{+\infty} n_{\mathrm{w}, 0} \cdot\left(\frac{r}{R_{\star}}\right)^{-2} \mathrm{d} l \\
& =\frac{\dot{M}}{4 \pi v_\mathrm{w}  \mu_\mathrm{i} m_\mathrm{p}} \int_{l_\mathrm{s} }^{+\infty} \frac{\mathrm{d} l}{d^2+l^2+2 d l \cos \left(\phi-\phi_{\mathrm{o}}\right) \sin \theta_{\mathrm{o}}} \\
& =\frac{\dot{M}}{4 \pi v_\mathrm{w} \mu_\mathrm{i} m_\mathrm{p} d} \int_{\tilde{l}_\mathrm{s} }^{+\infty} \frac{\mathrm{d} \tilde{l}}{1+\tilde{l}^2+2\tilde{l} \cos \left(\phi-\phi_{\mathrm{o}}\right) \sin \theta_{\mathrm{o}}} \\
& =\mathrm{DM}_{\mathrm{w,0}}\cdot\left[F(\phi, \tilde{l}=+\infty)-F\left(\phi, \tilde{l}=\tilde{l}_\mathrm{s}\right)\right],
\end{aligned}
\end{equation}
where $\tilde{l}=l/d$ is the dimensionless distance. The DM from the stellar wind depends on the $\mathrm{DM}_{\mathrm{w,0}}$ determined by the stellar wind and binary parameters
\begin{equation}
\begin{aligned}
\mathrm{DM_{w, 0}}&=   \frac{\dot{M}}{4 \pi v_\mathrm{w} \mu_\mathrm{i} m_\mathrm{p} a} \\
&\approx  1.1  ~\mathrm{p c~cm}^{-3}v_\mathrm{w,8}^{-1}\left(\frac{\dot{M}}{10^{-8} ~M_\odot \mathrm{yr}^{-1}}\right)\left(\frac{P_{\mathrm{orb}}}{1 ~\mathrm{yr}}\right)^{-\frac{2}{3}}\left(\frac{M_\star}{30~M_\odot}\right)^{-\frac{1}{3}},
\end{aligned}
\end{equation}
and the orbital modulation function 
\begin{equation}
\begin{aligned}
& F(\phi,\tilde{l})=\frac{1+e \cos \phi}{1-e^2} \frac{2 \arctan \left[\frac{f(\phi,\tilde{l})}{g(\phi)}\right]}{g(\phi)} \\
& g(\phi)=\left(3+\cos 2 \theta_0-2 \cos [2(\phi-\phi_{\mathrm{o}})] \sin ^2 \theta_{\mathrm{o}}\right)^{\frac{1}{2}}. \\
& f(\phi,\tilde{l})=2\left(\tilde{l}+\cos \left(\phi-\phi_{\mathrm{o}}\right) \sin \theta_{\mathrm{o}}\right)
\end{aligned}
\end{equation}
The magnetar wind cavity size $\tilde{l}_\mathrm{s}$ along the LoS (see panel (b) in Figure \ref{fig1}) can be obtained from the geometry of the bow shock as \citep{Canto1996}
\begin{equation}
\begin{aligned}
& l_{\mathrm{s}}=d \sin \theta_{\mathrm{s}} \csc \left(\theta_{\mathrm{s}}+\theta_{\mathrm{p}}\right) \\
& \theta_{\mathrm{s}}=\left\{\frac{15}{2}\left[\sqrt{1+\frac{4}{5} \eta\left(1-\theta_{\mathrm{p}} \cot \theta_{\mathrm{p}}\right)}-1\right]\right\}^{1 / 2}, \\
& \theta_{\mathrm{p}}=\pi-\arccos \left(\boldsymbol{e}_{\mathrm{mag}} \cdot \boldsymbol{e}_{\mathrm{obs}}\right)
\end{aligned}
\end{equation}
where $\eta=L_{\mathrm{sd}} / c/\dot{M} v_{\mathrm{w}}$ is the momentum rate ratio of the magnetar and the stellar wind. 

The temperature distribution of the adiabatically cooling stellar wind is \citep{Kochanek1993}
\begin{equation}
T_{\mathrm{w}}(r)=T_{\star}\left(\frac{r}{R_{\star}}\right)^{-\beta},
\end{equation}
where $T_{\star}$ is the effective temperature at the surface of the star. The power-law index $\beta\sim2 / 3-4 / 3$ depends on the adiabatic index of the stellar wind, and the value is set to $2/3$ in this work. The free-free absorption coefficient at frequency $\nu$ is $\alpha_\nu\simeq0.018 ~\mathrm{cm}^{-1}Z^2 n_{\mathrm{e}} n_{\mathrm{i}} T^{-3 / 2} \nu^{-2} \bar{g}_{\mathrm{ff}}$, where $\bar{g}_{\mathrm{ff}}$ is the Gaunt factor and $Z$ is the atomic number of the ions. The free–free optical depth of the stellar wind is
\begin{equation}
\begin{aligned}
\tau_{\mathrm{ff}}(\phi) &= \int_{l_\mathrm{s} }^{+\infty} \alpha_\nu(l)\mathrm{d} l \\
& \approx 0.018Z^2 T_{\star}^{-3 / 2}\nu^{-2}R_{\star}^{-1}\left(\frac{\dot{M}}{4 \pi v_\mathrm{w}  \mu_\mathrm{i} m_\mathrm{p}}\right)^2 \int_{l_\mathrm{s} }^{+\infty} \frac{\mathrm{d} l}{\left(d^2+l^2+2 d l \cos \left(\phi-\phi_{\mathrm{o}}\right) \sin \theta_{\mathrm{o}}\right)^{3/2}} \\
& =\tau_{\mathrm{ff,0}}\cdot\left[G(\phi, \tilde{l}=+\infty)-G\left(\phi, \tilde{l}=\tilde{l}_\mathrm{s}\right)\right].
\end{aligned}
\end{equation}
The free–free optical depth from the stellar wind is
\begin{equation}
\begin{aligned}
\tau_{\mathrm{ff,0}}&\approx0.018Z^2 T_{\star}^{-3 / 2}\nu^{-2}R_{\star}^{-1}\left(\frac{\dot{M}}{4 \pi v_\mathrm{w}  \mu_\mathrm{i} m_\mathrm{p}a}\right)^2 \\
&\approx  0.01T_{\star,5}^{-3 / 2}\nu_{9}^{-2}v_\mathrm{w,8}^{-2}\left(\frac{\dot{M}}{10^{-8} ~M_\odot \mathrm{yr}^{-1}}\right)^2\left(\frac{P}{1 ~\mathrm{yr}}\right)^{-\frac{4}{3}}\left(\frac{M_\star}{30~M_\odot}\right)^{-\frac{2}{3}},
\end{aligned}
\end{equation}
The orbital modulation function for the free–free optical depth is
\begin{equation}
G(\phi,\tilde{l})=\left(\frac{1+e \cos \phi}{1-e^2}\right)^2 \frac{4(\tilde{l}+\cos \left(\phi-\phi_{\mathrm{o}}\right) \sin \theta_{\mathrm{o}})}{\tilde{r}(\phi,\tilde{l})\left[3+2\cos2\theta_{\mathrm{o}}\cos^2 \left(\phi-\phi_{\mathrm{o}}\right)-2\cos 2\left(\phi-\phi_{\mathrm{o}}\right)\right]}, 
\end{equation}
where $\tilde{r}$ is given in Equation (\ref{eq:r}). Here, we ignore the effects of orbital geometry and estimate the maximum DM that the stellar wind can contribute, determined by $\tau_{\mathrm{ff,0}}=1$
\begin{equation}\label{DMwind}
\mathrm{DM}_{\tau_{\mathrm{ff,0}}=1}\approx 11.3~\mathrm{p c~cm}^{-3}T_{\star,5}^{3/4}\nu_9 \left(\frac{R_\star}{10~R_\odot}\right)^{\frac{1}{2}}.    
\end{equation}
%For typical values, the DM variation rate is $-12$ pc cm$^{-3}$ yr$^{-1}$ is the ejected mass is about 3.5$M_{\odot}$.

%The DM evolution during the Sedov–Taylor phase depends on the ionization fraction $f$ of the ambient medium
%\begin{equation}
%    	\frac{d{\rm DM}_{\rm SNR}}{dt} =
%	 0.72 (1-2.7f)
%	E_{51}^{1/5}
%	n_0^{4/5}
%	t_{1000\,\rm yr}^{-3/5}\,{\rm pc\,cm^{-3}}.
%    \nonumber
%    \\
%\end{equation}
%For $f<$ 0.37, the DM will increase with time. Because more ambient media are swept into the shock, they will be ionized, which will lead to the DM increasing in the SNR.

\subsection{DM and RM contributed by the SNR}\label{sec_SNR}
SNRs are also not a negligible component of the FRB surronding environment. This is particularly true for sources with a bright PRS, as the PRS emission strongly suggests the presence of a young SNR. Assuming that the ejecta of a supernova (SN) are ionized with an ionization fraction $f_e$, the DM contributed by a young supernova remnant (SNR) evolves with time as follows \citep{Piro2016,WangP2025}
\begin{equation}\label{eq:SNDM}
{\rm DM_{SNR}} = 26 \left(\frac{f_e}{0.1}\right) \left(\frac{E}{10^{51} \rm erg}\right)^{-1} \left(\frac{M_{\rm ej}}{1M_\odot}\right)^{2}
\left(\frac{t}{10 \rm yr}\right)^{-2}\,{\rm pc\,cm^{-3}},
\end{equation}
where $f_e$ is the ionized ratio of the SN ejecta in the free-expansion phase, $E$ is the kinetic energy, and $M_{\rm ej}$ is the ejecta mass. 

For a strong non-relativistic shock with an adiabatic index of $5/3$, the post-shock temperature is estimated as
\begin{equation}
	T_{\text{sh}} = \frac{3 \mu m_p v_{\text{sh}}^2}{16 k_{\text{B}}} \simeq 1.4 \times 10^7 \left( \frac{\mu}{0.61} \right) \left( \frac{v_{\text{sh}}}{10^3 \text{ km s}^{-1}} \right)^2 \text{ K},
\end{equation}
where $v_{\text{sh}}$ is the shock velocity. Considering the temperature of the shocked region is much higher than the typical temperature of the unshocked ejecta, e.g., $T_{\text{ej}}\sim 10^{4}$ K, we therefore neglect free–free absorption in the shocked region and consider only absorption by the cooler ionized ejecta in the following calculations. To estimate the effect of free-free absorption associated with the supernova ejecta, we convert the ejecta contribution to the dispersion measure into an emission measure. Assuming a homologously expanding, approximately uniform ionized ejecta, the characteristic ejecta radius is $R_{\rm ej}=v_{\rm ej}t$, where $v_{\rm ej}\simeq (2E/M_{\rm ej})^{1/2}$. The emission measure can then be approximated as ${\rm EM}_{\rm SNR}\simeq {\rm DM}_{\rm SNR}^{2}/R_{\rm ej}$. Combining this with equation~(\ref{eq:SNDM}), we obtain
\begin{equation}\label{eq:SNEM}
{\rm EM_{SNR}} \simeq 6.6\times10^{3}
\left(\frac{f_e}{0.1}\right)^2
\left(\frac{E}{10^{51} \rm erg}\right)^{-5/2}
\left(\frac{M_{\rm ej}}{1M_\odot}\right)^{9/2}
\left(\frac{t}{10 \rm yr}\right)^{-5}
{\rm pc\ cm^{-6}} .
\end{equation}
The corresponding free--free optical depth at radio frequencies is
\begin{equation}\label{eq:SNtau}
\tau_{\rm ff} \simeq 2.2\times10^{-3}
\left(\frac{T_e}{10^4 \rm K}\right)^{-1.35}
\left(\frac{\nu}{1 \rm GHz}\right)^{-2.1}
\left(\frac{f_e}{0.1}\right)^2
\left(\frac{E}{10^{51} \rm erg}\right)^{-5/2}
\left(\frac{M_{\rm ej}}{1M_\odot}\right)^{9/2}
\left(\frac{t}{10 \rm yr}\right)^{-5} .
\end{equation}
Here $T_e$ is the electron temperature of the absorbing ionized gas, and $\nu$ is the rest-frame observing frequency. This result shows that the free--free opacity decreases much more rapidly than the ejecta DM, with $\tau_{\rm ff}\propto t^{-5}$. Therefore, even if young supernova ejecta can provide a non-negligible local DM contribution, the ejecta become transparent to GHz radio emission on a relatively short timescale for standard parameters.

For the RM, we consider the magnetic field in the SN ejecta, assuming it follows energy-equipartition. Specifically, a $\epsilon_{B}$ fraction of the shock energy is converted into magnetic energy. In this scenario, the RM contribution from the SNR in the free-expansion phase is given by \citep{Piro2018,WangP2025}:
\begin{equation}\label{eq:RM}
	|{\rm RM_{SNR}}| \approx  2.9\times10^5 \left(\frac{f_e}{0.1}\right) \left(\frac{\xi}{0.1}\right) \left(\frac{\epsilon_{B}}{0.1}\right)^{1/2}
	\left(\frac{E}{10^{51} \rm erg}\right)^{-5/4}  \left(\frac{M_{\rm ej}}{1M_\odot}\right)^{11/4} \left(\frac{t}{10 \rm yr}\right)^{-7/2}\,{\rm rad\,m^{-2}}, 
\end{equation}
with the $\xi= \langle B_{ \parallel} \rangle/ \langle B \rangle$ is a parameter depending on the magnetic field geometry, and $\epsilon_{B}$ represents how much of the shock energy goes to the magnetic field. With the typical parameters, the theoretical RM evolution from equation (\ref{eq:RM}) follows the similar decreasing trend observed in FRB 20121102A.%For typical values, the RM decreasing rate is about $-10^{4}$ rad cm$^{-2}$ yr$^{-1}$ at the age of ten years.

\section{The RM and DM evolution in unified model}\label{sec_result}
\subsection{Application to FRB 20190520B}
\begin{figure} 
	\centering
	\includegraphics[width=\textwidth]{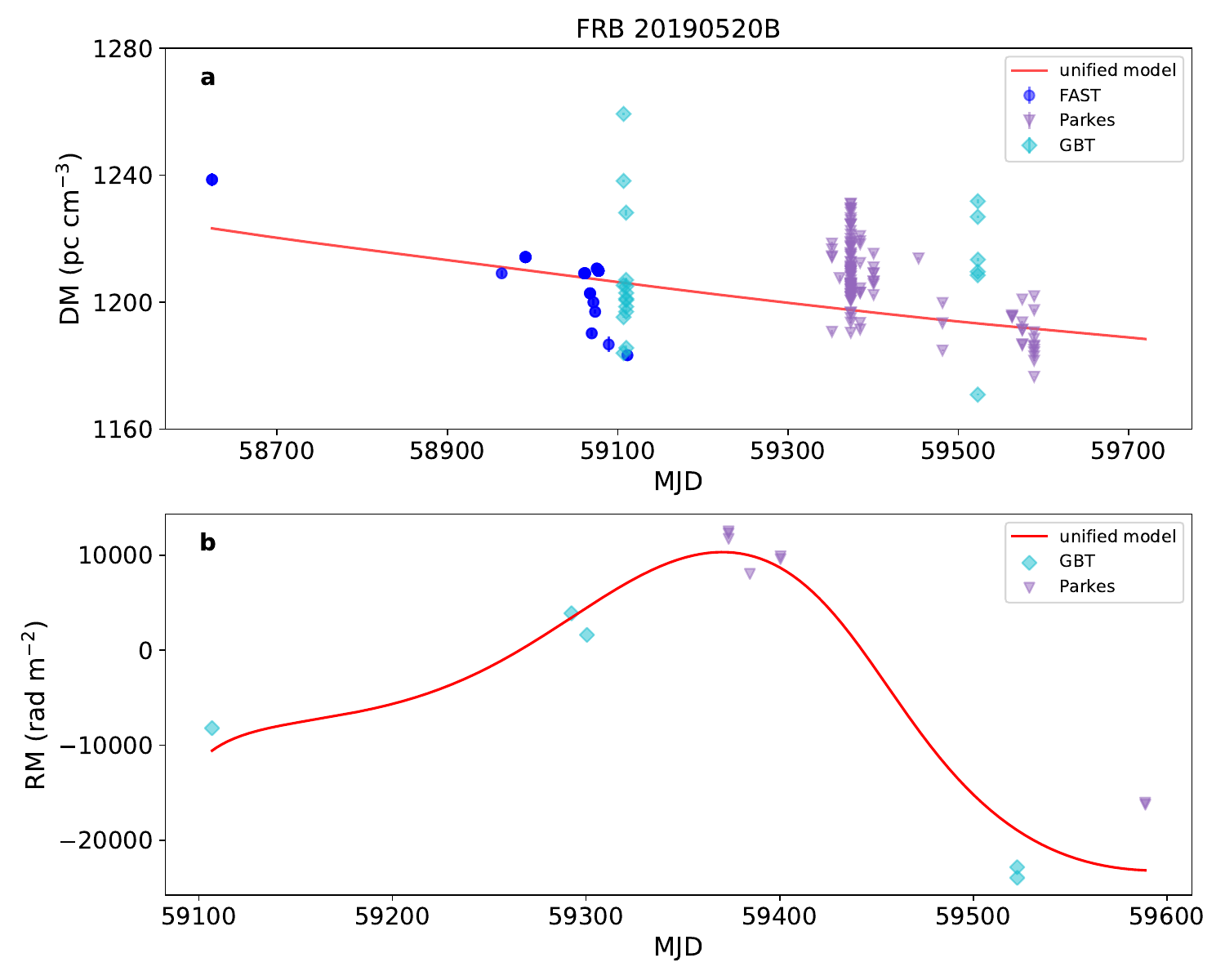}
	\caption{\textbf{The fits of DM and RM variations for FRB 20190520B with the unified model.} \textbf{a}, The temporal DM variation for FRB 20190520B. The red line shows the result of MCMC fit with the unified model. The green line is the DM contributed by SNR. The DM of stellar wind is shown in the inset figure. The observed DM value is shown as points. \textbf{b}, The temporal RM variation for FRB 20190520B. The red line shows the result of MCMC fit with the unified model, in which both the SNR (green line) and stellar wind (purple line) contribute significantly to the RM variation. The observed RM value is shown as points.
	}
	\label{fig2}
\end{figure}

We compile DM measurements of FRB 20190520B between 2019 and 2022 \citep{Niu2022,Anna-Thomas2023}, and find that it declines with a rate of about 11 pc cm$^{-3}$ year$^{-1}$ in four years, as shown in panel (a) of Figure \ref{fig2}. A similar trend is also found using more data \citep{Niu2026}. To quantify the significance of the decreasing trend, the null hypothesis method is used in the linear regression \citep{Wong2020,Jia2023}. In this method, the null hypothesis is that the slope of the linear fit is zero. The observational data is then used to test the validity of this null hypothesis. For the DM of FRB 20190520B, a t-value of 218.73 with 190 degrees of freedom was obtained, indicating that the confidence in the downward trend of the DM for FRB 20190520B is well above 10 $\sigma$. %\textbf{A t-test was used to assess the significance of this linear hypothesis for FRB 20190520B, obtaining a t-value of 218.73 with 190 degrees of freedom. This indicates that the confidence level for the declination trend in the DM of FRB 20190520B is well above 10 $\sigma$.} 

We consider some possible astrophysical processes that may cause DM variations. The DMs contributed by the stellar flares from a companion and magnetar flares decrease rapidly, typically dozens of days \citep{Yang2023,Xiao2025}. Obviously, they cannot generate the observed four-year DM decline. Below, we discuss the stellar winds from massive companions and SNRs.
We calculate the long-term DM evolution caused by the stellar winds in section \ref{sec_wind}. 
The maximal DM contributed by stellar winds is constrained by $\tau_{\mathrm{ff,0}}<1$, where $\tau_{\mathrm{ff,0}}$ is the optical depth of the free-free absorption for stellar winds.
After ignoring the effect of orbital geometry, the maximal DM contributed by the stellar wind is about a few dozens (equation (\ref{DMwind})), which is well below the decreased amount of FRB 20190520B in four years.

Therefore, the secular DM decline of FRB 20190520B cannot be caused by the stellar wind of a massive companion, but is consistent with a young SNR. For typical values of equation (\ref{eq:SNDM}), the DM decreasing rate is about $-11$ pc cm$^{-3}$ yr$^{-1}$ at $t=14$ years. So the rapid decline of DM is strong evidence of SNR. From the light curve and spectrum of the PRS, the age of the magnetar is found to be a few decades \citep{Zhao2021,Bhattacharya2024,Rahaman2025}, consistent with the above value. 

We use the unified model to fit the DM variation with the Markov Chain Monte Carlo (MCMC) method. The parameters in the MCMC fitting for FRB 20190520B include $P$, $B_0$, $\dot{M}$, $\Phi_0$, $\theta_o$, $\phi_o$, $i_m$ ,$\phi_m$ ,$a_{\mathrm{rm}}$ ,$a_{\mathrm{dm}}$, $t_{\mathrm{age}}$ and $\mathrm{DM}_0$. The first seven parameters are used to fit the impact of stellar wind, while the last four parameters are used to fit the influence of the SNR. Since the parameters of the SNR model are coupled together, these physical quantities cannot be directly fitted in practice \citep{Hilmarsson2021}. For convenience in fitting, $a_{\rm fit}$ is introduced to directly perform the MCMC fitting
\begin{equation}\label{eq_fit}
    y = a_{\rm fit}(t+t_{\rm age})^{-\alpha}+b,
\end{equation}
where $t$ is measured from the first detection of the FRB source, and $t_{\rm age}$ represents the age of the FRB source at the time of its first detection \citep{Zhao2021}. For FRB 20190520B, which was first discovered in 2019, an inferred value of $t_{\rm age}\sim20$ yr therefore means that the SNR was formed approximately 20 years before 2019, i.e., around 1999. For the DM contributed by the SNR, $a_{\rm fit}=a_{\mathrm{dm}}=26 f_{e,-1}E_{51}^{-1}M_0^{2}$, $\alpha=-2$ and $b$ represents DM$_0$ contributed by other terms \citep{WangP2025}. For the RM contributed by the SNR, $a_{\rm fit}=a_{\mathrm{rm}}=2.9\times 10^{6} f_{e,-1} \xi \epsilon_{B,-1}^{1/2} E_{51}^{-5/4} M_{0}^{11/4}$, $\alpha=-7/2$ and $b$ represents RM$_0$ contributed by other terms. Considering that the RM contribution from other terms is much smaller than that of SNR, RM$_0$ is set to 0 when fitting the RM for both sources. In the MCMC fitting process, to ensure that the physical parameters of the unified model remain within a reasonable range, the parameter range of $a_{\rm rm}$ was set from $10$ to $10^{9}$ and $a_{\rm dm}$ was set from $1$ to $10^{5}$. It is important to note that, due to the coupling of several parameters, only selected parameters were included in the MCMC fitting, whereas the remaining ones were fixed to reasonable values, e.g. the eccentricity $e=0.5$. For the physical parameters of the SNR, the kinetic energy $E$ for FRB 20190520B is set to $10^{51}$ erg, and the ejcta mass $M$ is set to $1.83\ M_{\odot}$.

\begin{figure}
	\centering
	\includegraphics[width=\columnwidth]{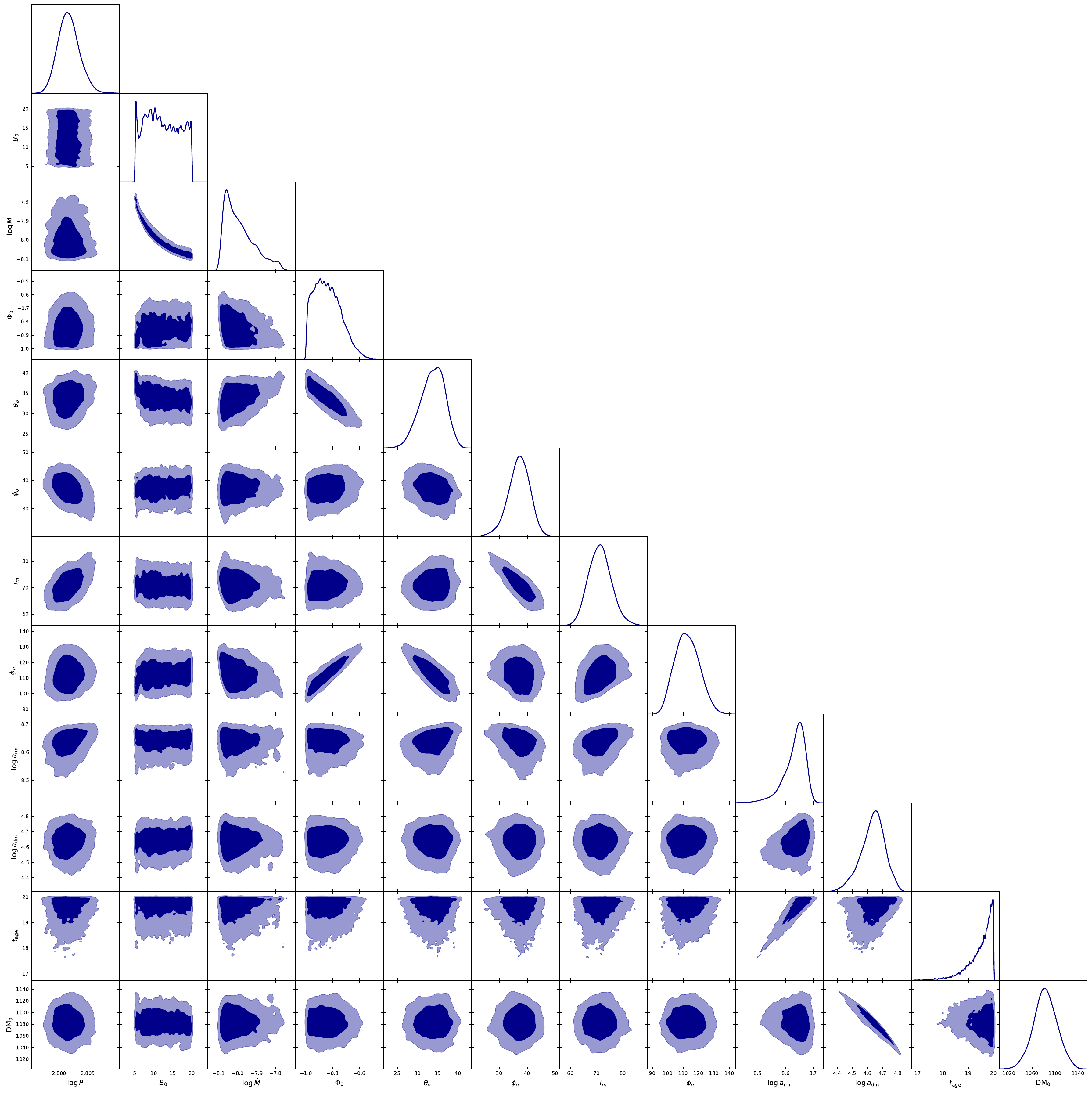}
	\caption{\textbf{Two-dimension posterior corner plot for the parameters of the unified model for FRB 20190520B.} The histograms indicate the posterior probability of each parameter. The plots show the explored parameter space, with $1\sigma$ and $2\sigma$ contours. }\label{fig_MCMC_190520}
\end{figure}

The best fit of DM is shown as the solid line in panel (a) of Figure \ref{fig2}. The corner plot is given in Figure \ref{fig_MCMC_190520}. The best-fit parameters are given in Table \ref{table1}. The fit of FRB 20190520B yields robust results. The RM of FRB 20190520B was highly variable. It shows the largest reversal in the Universe. The RM variation can be naturally reproduced in the magnetar-massive star binary \citep{Wang2022,Anna-Thomas2023}. The radio waves traveling through the stellar wind of a massive companion star will cause the observed RM variation. The unified model is used to fit the RM variation, as shown in panel (b) of Figure \ref{fig2}. This model can naturally explain the DM and RM variations of this FRB source.
In addition, the free--free optical depth associated with the SNR is only 
\(\tau_{\rm ff}=0.03\), indicating that free--free absorption by the SNR 
does not significantly attenuate the FRB emission. To further verify whether the additional complexity of the unified model is truly justified by the data. We calculate the Bayesian Information Criterion (BIC) for the three model: SNR model, binary model and our unified model. For FRB 20190520B, the BIC of SNR model is 3023.88, the BIC of Binary model is 363.12 and the BIC of our unified model is 138.28. This results shows that the additional complexity of the unified model is truly justified.

%where $\mu_e$ is the mean molecular weight per electron, $E$ is the kinetic energy, $M_{\rm ej}$ is the ejecta mass, and $K$ is wind mass loading parameter. \textbf{For typical values, the DM decreasing rate is about $-11$ pc cm$^{-3}$ yr$^{-1}$ at $t=14.5$ years. The So the rapid decline of DM is a strong evidence of SNR. From the light curve and spectrum of the PRS, the age of magentar is found to be *** years, consistent with the above value. We use the unified model to fit the DM variation with Markov Chain Monte Carlo (MCMC) code, shown as the solid line in panel (a) of Figure \ref{fig2}. The best-fit parameters are given in Table S\ref{table1}}.

%\begin{equation}
%\frac{d{\rm DM}_{\rm SNR}}{dt}  \approx -92 \left(\frac{f}{0.1}\right) \left(\frac{M_{\rm ej}}{10~M_{\odot}}\right) \left({\frac{t}{10\rm yr}}\right)^{-3}\,{\rm pc\,cm^{-3}\,yr^{-1}}, 
%\end{equation}
%where $f$ is the ionized fraction of the ejecta, and $M_{\rm ej}$ is the ejecta mass.
%For typical values, the DM variation rate is $-11$ pc cm$^{-3}$ yr$^{-1}$ for the ejected mass of 1$M_{\odot}$. So the rapid decline of DM is a strong evidence of SNR. 

\subsection{Application to FRB 20121102A}
Unlike FRB 20190520B, the RM of FRB 20121102A decreases monotonically, i.e., from $10^5~\mathrm{rad\,m^{-2}}$ to $3\times 10^4~\mathrm{rad\,m^{-2}}$ at the year of 2023 \citep{Michilli2018,Hilmarsson2021,WangP2025}. This descending trend is consistent with the prediction of a young SNR.

\begin{figure} 
	\centering
	\includegraphics[width=\textwidth]{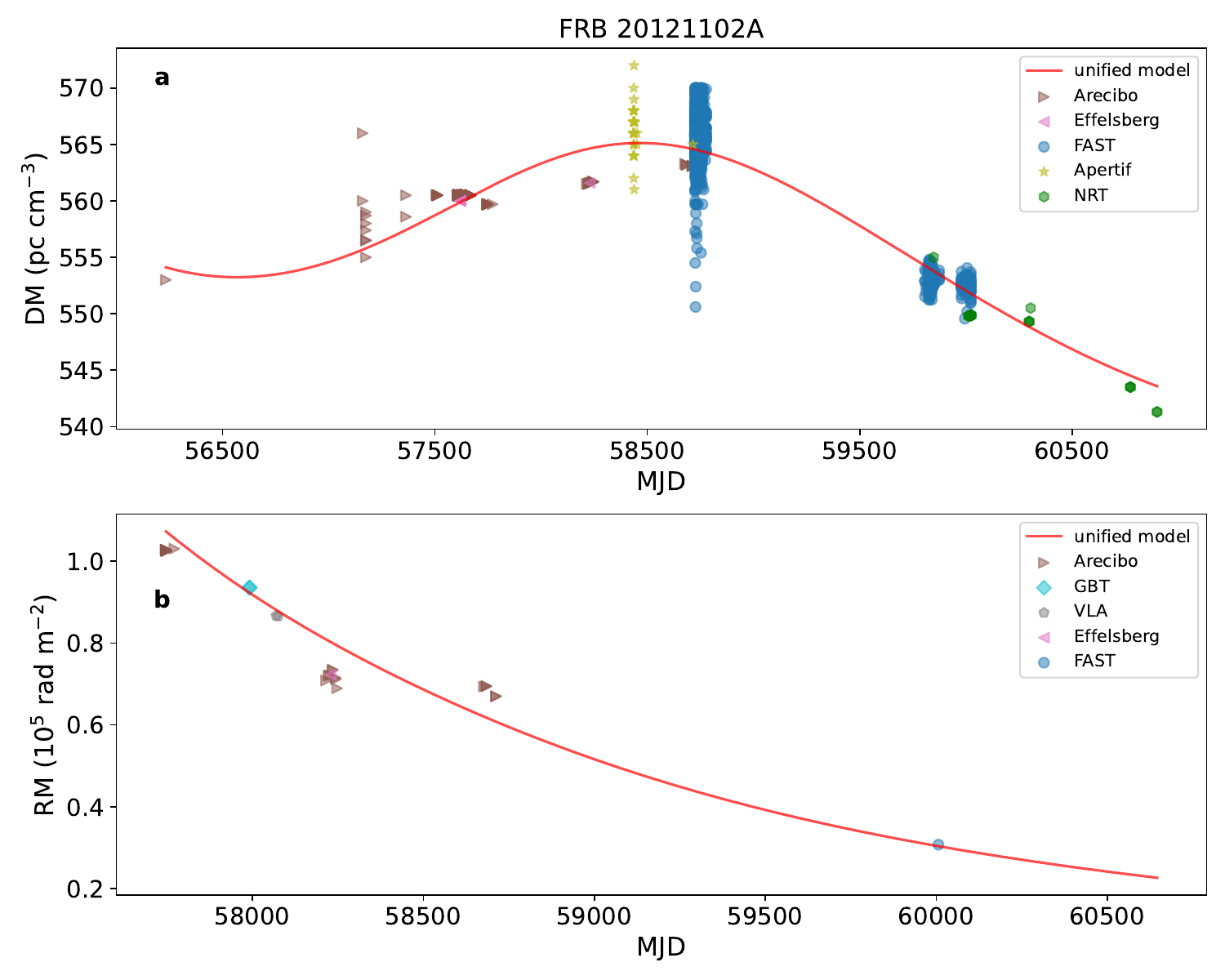}
	\caption{\textbf{The fitting result for FRB 20121102A with the unified model.} \textbf{a}, The temporal DM variation of FRB 20121102A. The red line shows the result of MCMC fit with the unified model, in which both the SNR (green line) and stellar wind (purple line) contribute significantly to the DM variation. The observed DM value is shown as points. \textbf{b}, The red line shows the RM fit with the unified model. The green line shows the RM contribution by the SNR. The points represent the observed RM.  
	}
	\label{fig3}
\end{figure}

However, the DM of this source increases during eight years since its discovery \citep{Hessels2019,Oostrum2020,Li2021}, and then declines \citep{WangP2025,Snelders2025}, as shown in 
Figure \ref{fig3}. With use of the null hypothesis method mentioned above, a t-value of 109.290 with 2218 degrees of freedom was obtained, showing that the confidence in the upward trend of the DM for FRB 20121102A before MJD 58,500 is also well above 10$\sigma$. The increasing trend is consistent with that seen before \citep{Hessels2019,Oostrum2020,Li2021}. However, a flat DM behaviors is found in this period \citep{WangP2025}. The main reason is that they omit the DM measurement in the year 2012. The long-term trend relies heavily on earlier DM measurements \citep{Li2021}. Therefore, the increasing trend of DM for FRB 20121102A before MJD 58500 is robust. From equation (\ref{eq:SNDM}), the DM contributed by an SNR will decrease in the free expansion phase, and then increase in the Sedov-Taylor phase due to the ionization of the ambient medium \citep{Piro2016,Yang2017}. Therefore, a young SNR cannot explain the DM and RM variation simultaneously. Our unified model can naturally explain its DM and RM variations. When the magnetar approaches the massive star, the DM will increase as radio bursts interacting more dense stellar wind. As the magnetar departs from the periastron, the DM will decrease. This is consistent with the DM variation of FRB 20121102A. 

\begin{figure}
	\centering
	\includegraphics[width=\columnwidth]{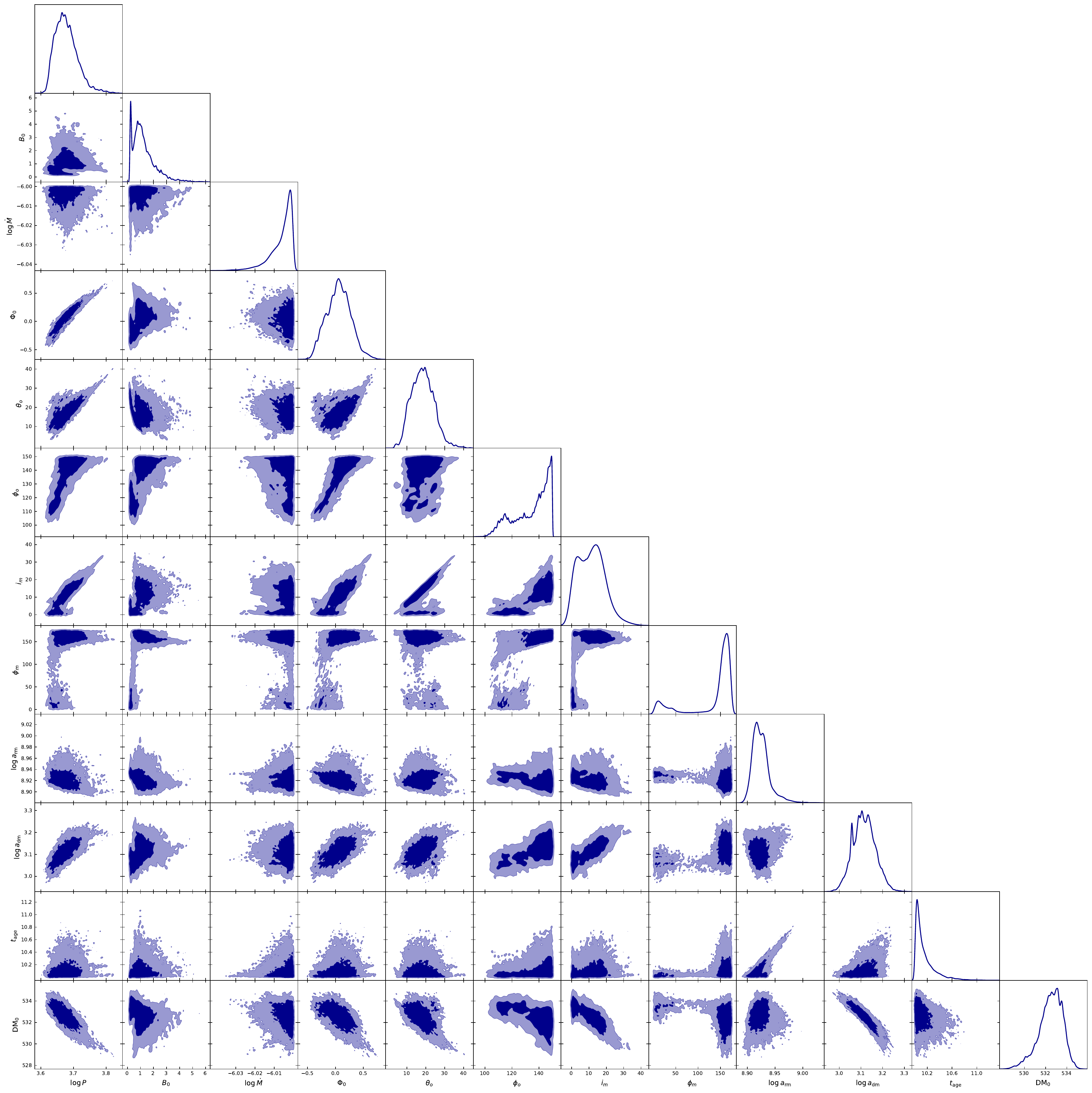}
	\caption{\textbf{Two-dimension posterior corner plot for the parameters of the unified model for FRB 20121102A.} The histograms indicate the posterior probability of each parameter. The plots show the explored parameter space, with $1\sigma$ and $2\sigma$ contours. }\label{fig_MCMC_121102}
\end{figure}

We use the same MCMC method to fit the DM and RM variations. In Figure \ref{fig3}, we show the fitting result the DM and RM variations of FRB 20121102A with the unified model. The corner plot is given in Figure \ref{fig_MCMC_121102}. The best-fit parameters are shown in Table 1. The age of new-born magnetar is about ten years, which is also consistent with the age derived from the associated PRS \citep{Beloborodov2017,Margalit2018,Hilmarsson2021,Zhao2021}. The derived period is about 5700 days, which is larger the possible active period of 157 days \citep{Rajwade121102,Cruces2021}. This supports that the active period is not caused by binary orbital motion, similar as that of FRB 20180918B \citep{CHIME/FRB2020,Pastor-Marazuela2021}. In addition, the free--free optical depth associated with the SNR is only 
\(\tau_{\rm ff}=4.22\times10^{-4}\), indicating that free--free absorption by the SNR 
does not significantly attenuate the FRB emission. For FRB 20121102A, the BIC of SNR model is 1856.30, the BIC of binary model is 2525.77 and the BIC of our unified model is 689.43. This results also shows that the additional complexity of the unified model is truly justified.

These two active FRBs are associated with bright PRSs, which can be attributed to synchrotron emission from relativistic charged particles of an MWN. Energy injection into the nebula surrounding the central magnetar can be the rotational
energy for a young magnetar \citep{Kashiyama2017,Bhattacharya2024}, and the release of magnetic
energy from magnetar interior \citep{Beloborodov2017,Margalit2018,Zhao2021}. Decade-old magnetars are required to account for the
observed PRSs for these two FRB sources \citep{Margalit2018,Yang2019,Hilmarsson2021,Zhao2021,Bhattacharya2024,Rahaman2025}, which is consistent with the age estimation from their DM and RM variations, as shown in Table \ref{table1}. The detailed modeling the light curves and spectra of PRSs are given in \cite{Zhao2026}.

\subsection{Application to other four active FRB sources}
To test the generality of the unified model, we applied it to other three active FRB sources associated with PRSs, including FRB 20201124A, FRB 20190417A and FRB 20240114A. 
From long-term monitoring, their DM shows stochastic fluctuation \citep{Xu2022,ZhangJ2025,Moroianu2025}. No secular DM evolution is found. This indicates that the contribution of an SNR is negligible, possibly due to a relative old age of magnetars or a low ionization fraction in SN ejecta. We mainly focus on the RM variation caused by massive companion stars.
The RM variation of FRB 20201124A is complex. It shows oscillation in short timescale, followed by a
constant RM phase \citep{Xu2022}. After subtracting the RM value of the constant phase, the RM contributed the source vicinity
shows sign reversal \citep{Wang2022}. The RM variation can be well reproduced in the magnetar-massive star binary \citep{Wang2022}. Subsequent monitoring this source found that periodic variation of RM from more than 3000 bursts collected by the Five-hundred-meter Aperture Spherical radio Telescope (FAST) \citep{XuJ2025}, which is direct evidence for the binary model. 

\begin{figure} 
	\centering
	\includegraphics[width=\textwidth]{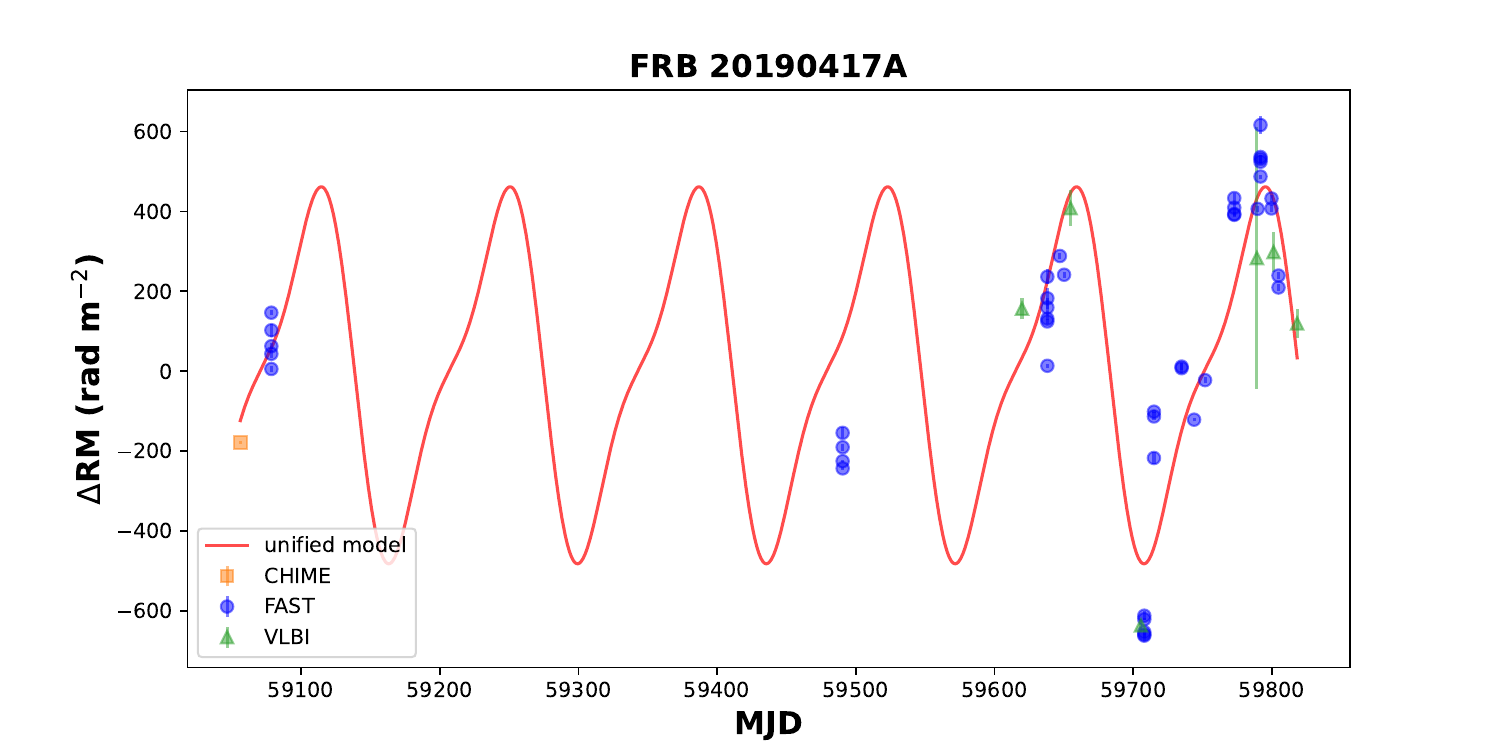}
	\caption{\textbf{The RM fit of FRB 20190417A using the unified model.} The RM variation is mainly contributed by the stellar wind. The points represent the observed RM variation $\Delta \rm RM_{obs}=RM_{obs}-RM_0$, where RM$_0$ is constant RM contributed by other terms. 
    }
	\label{fig4}
\end{figure}

FRB 20190417A shows large RM variation from about 4000 to 5000 rad m$^{-2}$ \citep{Feng2025,Moroianu2025}. Figure \ref{fig4} shows the RM fit using the unified model. The fitted parameters are shown in Table \ref{table1}. It should be noted that, due to the limited observations, the short time span of the events, and possible parameters coupling, the fitted physical parameters are not unique. Tighter constraints on these parameters will require long-term observations. 

The RM of FRB 20240114A also shows secular variation \citep{Xie2025}. Long-term monitoring is required to test a possible quasi-periodic RM evolution. %FAST has detected more than ten thousands bursts in 216-day observation with a peak rate of 729 hr$^{-1}$ \citep{ZhangJ2025}. 
It should be noted that the stellar wind or decretion disk of massive stars are clumpy \citep{Puls2006,Zhao2023}. For example, the pulsar binary system PSR B1259-63/LS 2883 with 3.5 year orbital period shows distinct RM values at similar orbital phases \citep{Johnston1996}, indicating a clumpy wind or disk \citep{Melatos1995}. So a strict period is hard to derived only from RM variation. 

These three FRB sources are also associated with PRSs \citep{Bruni2024,Bruni2024b,Moroianu2025,ZhangX2025}, which can be explained by the MWN model \citep{Bruni2024,Bruni2024b,Moroianu2025,Zhao2026}. The luminosity is about one order of magnitude lower than the first two PRSs. Theoretically, the luminosity of MWNs is scaled as $\propto t^{-2}$ \citep{Margalit2018}. If the parameters (mass, spin period, magnetic field) of the newborn magnetars are similar, the low luminosity may be due to a relatively old magnetar. As the magnetar becomes older, the energy injection into the nebula from the rotational energy decreases rapidly. Meanwhile, due to a short magnetic active timescale \citep{Beloborodov2016}, the energy injection by the magnetic energy from magnetar interior will decay, which produces a low-luminosity PRS. As shown in equations (\ref{eq:SNDM}) and (\ref{eq:RM}), the DM and RM contributed by SNRs decrease rapidly with time \citep{Piro2018}. The stochastic variations of DM and low RM of these three active FRBs support that the age of a central magnetar is relative old, compared to those of FRB 20121102A and FRB 20190520B.

For FRB 20220529A, the RM flare and possible temporal RM period support the binary motion as the origin of RM variation. More recently, a gradual decline in DM for this FRB source is found by CHIME/FRB \citep{Pandhi2026}, implying an expanding SNR. Therefore, the temporal RM and DM variations show strong evidence of a binary embedded in an SNR.

\section{Conclusions and Discussion}\label{Conclusion and Discussion}
These repeating FRBs have very high burst rates. The massive companion can play a significant role for boosting the burst triggers. 
Known magnetars in the Milky Way and nearby galaxies are in isolated systems, and none of them have a known companion star. Population syntheses reveal that a few percent of magnetars can be formed in binary systems, most with a massive star companion \citep{ZhangG2020,ZhangB2025}. %These active repeaters haveextremely high burst triggering rate. 
The massive companion may play a major role. The massive star companion not only provides RM and DM variations, but also triggers FRBs. One hypothesis is that the stellar wind could be accreted into the magnetar magnetosphere, potentially causing instabilities that lead to the triggering of FRBs \citep{Lan2024,ZhangB2025}. Alternatively, it is hypothesized that the companion wind might disturb the magnetosphere, initiating the propagation of Alfv\'{e}n waves downward. These waves may induce magnetic instability or collide with Alfv\'{e}n waves launched from the magnetar surface, which may trigger FRBs \citep{ZhangB2025}.

In a summary, we have demonstrated that active FRBs associated with PRSs can be understood within a unified model in which the sources are young magnetars embedded in SNRs in massive binary systems.
The diversity in the observed DM and RM temporal variations arises from different properties of SNR, binary parameters and viewing geometry.
As the age of a magnetar gets older, the luminosity of PRS will fade due to the magnetar spin-down or internal magnetic field decay, which can naturally explain the low luminosity of PRSs for some active FRBs. There is a possible evolutionary connection between these active FRBs, i.e., the younger systems hosting the more luminous PRSs and larger RMs, also supported by the empirical correlation between PRS luminosity
and RM magnitude \citep{Yang2020a}, and the positive correlation between RM and its standard deviation \citep{Feng2022}. In addition, the massive star companion can trigger FRBs with high burst rate. With long-term monitoring of active repeaters, we expect quasi-periodic DM or RM evolution if the stellar wind plays a dominate role. 

\section*{acknowledgments}
We thank an anonymous referee for helpful comments. This work was supported by the National Natural Science Foundation of China (grant Nos. 12494575, 12273009 and 12393812), and the National SKA Program of China (grant Nos. 2022SKA0130100 and 2020SKA0120302).  Y.F. is supported by National Natural Science Foundation of China grant No. 12203045, and by the Leading Innovation and Entrepreneurship Team of Zhejiang Province of China grant No. 2023R01008. This work made use of data from the Five-hundred-meter Aperture Spherical radio Telescope (FAST), a Chinese national mega-science facility built and operated by the National Astronomical Observatories, Chinese Academy of Sciences.

\bibliography{ms}{}
\bibliographystyle{aasjournal}

\clearpage

\begin{table}
    \centering
    \caption{The fitting results for the unified model}
    \resizebox{\textwidth}{!}{%
    \begin{tabular}{c c|c c c c}
    \hline
    \hline
    parameter & Symbol & FRB 20190520B & FRB 20121102A & FRB 20190417A \\ 
    \hline
     Age of magnetar & $t_{\mathrm{age}}\ \mathrm{(year)}$ & $19.47^{+0.53}_{-0.10}$  & $10.15^{+0.16}_{-0.16}$ & - \\ 
    Fitting parameter for RM & $a_{\mathrm{rm}}$ & $4.29^{+0.48}_{-0.19}\times10^{8}$ & $8.41^{+0.02}_{-0.32}\times10^{8}$ &  - \\ 
    Fitting parameter for DM & $a_{\mathrm{dm}}$ & $4.35^{+0.91}_{-0.66}\times10^{4}$ & $1.30^{+0.15}_{-0.17}\times10^{3}$ &  - \\ 
    Magnetic field geometry ratio & $\rm \xi$& $0.1$ & $0.2$ & - \\ 
    Kinetic energy of SNR & $ E\ (\rm erg)$& $10^{51}$ & $10^{51}$ & - \\
    Mass of the SN ejcta & $M$ ($M_\odot$)& $1.83$ & $0.71$ & - \\
    Ionized ratio  & $f_e$& $0.5$ & $0.1$ & - \\
    Magnetic energy fraction & $\epsilon_B$& $3.16\times10^{-5}$ & $0.14$ & - \\
    \hline
    Orbital period & $ P\ (\rm day)$ & $785.55^{+2.37}_{-4.15}$  & $5696.92^{+236.58}_{-450.15}$ & $136.11^{+32.16}_{-2.76}$ \\ 
    Eccentricity & $e$ & $0.5$  & $0.5$ & $0.5$ \\ 
    Fitting parameter & $\Phi_0$ & $-0.86^{+0.07}_{-0.14}$ & $0.30^{+0.22}_{-0.22}$ & $-0.71^{+3.91}_{-0.79}$ \\ 
    \hline
    $\rm $ & $\mathrm{RM}_0$ $\rm (rad\ m^{-2})$ & - & - &  $4608.52^{+155.79}_{-28.16}$ \\ 
    $\rm $ & $\mathrm{DM}_0$ $\rm (pc\ cm^{-3})$ & $1081^{+21}_{-21}$ & $528.62^{+1.3}_{-0.78}$ & - \\ 
    \hline
    Surface magnetic field & $B_0$ $\rm (G)$ & $7.79^{+4.30}_{-4.30}$  & $1.26^{+0.2}_{-1.1}$ & $1.3^{+0.7}_{-1.179}$ \\ 
    Mass-loss rate & $\dot{M}$ ($M_\odot$) & $1.2^{+0.38}_{-0.08}\times10^{-8}$  & $9.86^{+0.04}_{-0.14}\times10^{-7}$ & $8\times10^{-10}$ \\ 
    \hline
    Inclination angle of observers & $\theta_{\mathrm{o}}$ $\rm (degree)$& $34.40^{+3.40}_{-2.60}$  & $18.30^{+6}_{-6}$ & $6.64^{+29.36}_{-6.64}$ \\ 
    True anomaly angle of observers & $\phi_o$ $\rm (degree)$ & $37.20^{+3.80}_{-3.20}$  & $134^{+20}_{-20}$ & $2.9^{+105.1}_{-2.9}$ \\ 
    \hline
    Magnetic inclination angle & $i_m$ $\rm (degree)$& $70.90^{+4.20}_{-4.20}$  & $11.10^{+6.60}_{-9.10}$ & $1.23^{+104.77}_{-1.23}$ \\ 
    True anomaly angle of magnetic inclination & $\phi_m$ $\rm (degree)$& $111.00^{+6.60}_{-8.60}$  & $134^{+40}_{-10}$ & $143.41^{+27.59}_{-72.41}$ \\ 
    \hline
    \end{tabular}%
    }
    \label{table1}
\end{table}



\end{document}